\documentclass[aps,prd,twocolumn,groupedaddress,amssymb,eqsecnum,showpacs,epsfig,nofootinbib]{revtex4}
\usepackage{graphicx}
\usepackage{bm}
\usepackage{dcolumn}
\usepackage{amsmath}
\usepackage{amsmath}
\usepackage{amssymb}
\usepackage{epsfig}
 \newcommand{\nn}{\nonumber}
\newcommand{\dd}{{\rm d}}

\begin{document}
\title{Explanation for the acceleration of the universe's expansion without dark energy}
\author{Dong-Biao Kang}
\email{billkang@itp.ac.cn}
\address{Institute of Theoretical Physics, Chinese Academy of Sciences, Beijing 100190, China}

\begin{abstract}
{ We have observed the acceleration of the expansion of
the universe. To explain this phenomenon, we usually introduce the
dark energy (DE) which has a negative pressure or we need to modify the
Einstein's equation to produce a term which is equivalent to the
dark energy. Are there other possibilities? Combining our previous works
of statistical mechanics of self-gravitating system with the derivation of van der waals
equation, we propose a different matter's equation of state (EoS) in this paper.
Then we find that if the matter's density is low enough, its pressure can be negative, which
means that it is the matter that drives the expansion's acceleration.
So here we will not need to add the DE to the universe. Our results
also predict that the universe finally tends to be dominated by an
approximate constant energy density, but its value can be smaller than DE.
The data of Supernova can not differentiate our model from
the standard model, but they may indicate some deviations from $\Lambda CDM$.}

\end{abstract}


\maketitle

\section{Introduction}

Under the assumption of isotropy and homogeneity of our Universe at
large scale structure, people made a successful explanation to the
cosmology accelerating expansion~\cite{SNIa} by introducing a new
component called DE, which constitutes of about $70\%$
energy of the universe. Up to now, the most successful view about
the DE is the $\Lambda CDM$ model, in which the energy density of DE
is constant. Although $\Lambda CDM$ model is consistent very well
with all observational data, it faces the fine tuning problem
\cite{finetunning}. Alternatively, plenty of other DE models have
also proposed
\cite{Copeland:2006wr,Caldwell:1997mh,Steinhardt:1999nw,Capozziello:2003tk},
but almost all of them solve the acceleration problem either by
introducing new degree of freedom or by modifying gravity, which are
challenges both in Cosmology and in Nuclear Physics.

However, do we really need the DE? Or do we need so much DE? We notice that commonly the matter's EoS comes from the classical thermodynamics without consideration of the gravity, has it been confirmed to be reasonable at the cosmological scale? We think that the answer may be not sure, and our latest studies have provided different understandings of the long-range statistical mechanics. In fact, if we apply the Boltzmann entropy
\begin{equation}
\label{BG}
S = - \int f \ln f \dd \tau
\end{equation}
into the self-gravitating system, we will always obtain the infinite mass and energy with the principle of maximum entropy \cite{galdyn08}, which is a serious problem. \cite{hk10} preliminarily studies the entropy form taken by \cite{white87} and proposes that a self-gravitating system's entropy is at a saddle point and not a maximum. \cite{kh11} completes the variation process of entropy and confirms that we can obtain an EoS which is different from the classical one but can solve the problem we mentioned above. In \cite{kang11} we compare this EoS with the van der waals equation and speculate that our EoS should be a requirement of long-range statistical mechanics, which is also shown in this paper.

In this paper, we think that the general relativity can also have similar effects on the matter's EoS at the cosmological scale. So in the next section we will show our findings and propose a reason for the universe's accelerating expansion, which will not need the DE. In section \ref{snIa}, we will use the data of supernova to shorten the range of the values of the parameters in our model. Some discussions and conclusions will be made in the final section. In this paper we set $c=1$ and use ``$0$'' to denote quantity's current value except $\Omega_{x0}$ below.
\section{Explanation for the acceleration}
According to the cosmological principle, the universe is homogeneous and isotropic. Commonly the matter can be treated as the ideal fluid whose energy-momentum tensor can be written as
\begin{equation}
T^{\mu\nu}=(\rho + p)U^{\mu}U^{\nu}+pg^{\mu\nu}.
\end{equation}
where $\rho$ is the energy density, $p$ is the pressure, and $U^{\mu}$ is the velocity four-vector. When we substitute it into the Einstein's equation, we can get the Friedmann equation:
\begin{equation}
\label{friedmann}
\ddot{a}=-4\pi G(\frac{\rho}{3}+p)a, \dot{a}^2+\kappa=\frac{8\pi G}{3}\rho a^2,
\end{equation}
where $a$ is the scale factor. The energy-momentum conservation law is
\begin{equation}
\label{conserve}
\dot{\rho}=-\frac{3\dot{a}}{a}(\rho+p),
\end{equation}
which also can be obtained from eq.(\ref{friedmann}). Combining the EoS $p=p(\rho)$ with eq.(\ref{friedmann}), we can get the solution of $a$, $p$ and $\rho$. After the inflation, the universe is assumed to be a gas whose pressure is \cite{liddle00}
\begin{equation}
\label{class-p}
p=\frac{v^2}{3}\rho,
\end{equation}
where $v^2$ is the mean-square velocity, $m$ is the mass of particle, and $T$ is the defined temperature. We usually use $p=0$ and $p=\rho/3$ to denote the matter ($v^2=\frac{3kT}{m}<<1$) and radiation's ($v=1$) EoS respectively, but by eq.(\ref{class-p}) we can not understand the nature of the DE whose pressure is negative, although we commonly treat that the DE does negative work to the universe.

Notice that eq.(\ref{class-p}) can be derived by two ways \cite{Pathria96}: we can use the formula of the classical statistical mechanics to calculate the matter and radiation's EoSs, both of which satisfy eq.(\ref{class-p}); it also can be obtained by the kinetics alone, which needs to consider the bombardment by the particles (rigid body) on the walls of the container to calculate the kinetic pressure. However, the classical statistical mechanics does not consider the effects of gravity. Besides, the wall of the container does not exist in the universe, and in fact we can treat that the wall is equivalent to a potential well with minus infinity at the boundary for the free particles, while the gas in the universe is in the gravitational potential well which is different from the wall. If we treat the universe as a thermodynamical system where the particles interact by gravity, will these two ways mentioned above still be available under the cosmological background where the effect of general relativity can not be neglected? We think that the answer is yes for the radiation, because the effect of gravity can be neglected compared with the electromagnetic force; but for the matter whose main component is the dark matter where the gravity plays the most important role, the answer may be not sure. We consider this problem by two sides. On the one hand, in \cite{kh11} by the method of statistical mechanics we have obtained an EoS which is used to describe the thermodynamical equilibrium state of a virialized self-gravitating system:
\begin{equation}
\label{EoS}
\rho=\beta p + \alpha p^{4/5},
\end{equation}
where $\beta$ and $\alpha$ are positive, and $\beta$ has been identified to $m/kT$ in \cite{kang11}. Notice that here we take the Newtonian gravity and the power index of the second term of eq.(\ref{EoS})'s right ride must be less than 1 to ensure of the finite mass and energy of the system. From eq.(\ref{EoS}) we can obtain $p=p(\rho)$, which can be approximately written as \cite{kang11}
\begin{equation}
\label{eos3}
p=\frac{kT\rho}{m}-\alpha(\frac{kT}{m})^{\frac{9}{5}}\rho^{\frac{4}{5}}.
\end{equation}
Of course, eq.(\ref{eos3}) is different from eq.(\ref{class-p}) even we let $v^2=3kT/m$, and some different results possibly can be obtained. On the other hand, we consider the van der waals equation:
\begin{equation}
\label{vander}
(p+\frac{n^2a}{V^2})(V-nb)=NkT,
\end{equation}
where $a$ and $b$ is used to describe the attractive and repulsive interactions among the molecules of gas respectively, and they are determined by the potential of the molecules' interactions by the method of cluster expansion in the canonical ensemble \cite[see some books of statistical mechanics such as][]{Pathria96}, so we emphasize that the existence of $a$ and $b$ is determined by the statistical mechanics. If the potential of the molecules' interactions has other forms, $a$ and $b$ may still appear in eq.(\ref{vander})but their values and other places may be changed. Especially when we set $V=Nm/\rho$ and $b=0$, eq.(\ref{vander}) becomes
\begin{equation}
p=\frac{kT}{m}\rho-a(\frac{n}{Nm})^2\rho^2,
\end{equation}
and is very analogous to eq.(\ref{eos3}), which makes us speculate that existence of $\alpha$ in eq.(\ref{eos3}) also is a requirement of possible statistical mechanics which can deal with systems with long-range interaction of the Newtonian gravity, and the change of index from $2$ to $4/5$ may be caused by the change from the interactions among molecules to the Newtonian gravity. Then what is the case of that the particles are in the gravitational field described by general relativity? The answer is that such theory still does not set up now, but we may do some guess of its results as the following: based on above analysis, we find that $\beta$ is related to the temperature, $\alpha$ is necessary to denote the effect of the long-range statistical mechanics, and neither of them will disappear if the gravity changes from the Newtonian to the general relativity; but the index $4/5$ in eq.(\ref{eos3}) may be changed in the general relativity. So we assume that at the cosmological scale the EoS of the matter may be written as
\begin{equation}
\label{eos2}
p_m=\frac{v^2}{3}\rho_m - \Omega_{x0}\rho_0^{1-t}\rho_m^t, t<1, \Omega_{x0}>0,
\end{equation}
where $\rho_0=\frac{3H_0^2}{8\pi G}$, $t$ is a constant, and $v$ and $\Omega_{x0}$ are about constants or very slowly changing functions of time at the matter dominated era. Here we temporally do not assume that $v^2$ is as small as its value in $\Lambda CDM$. Notice that the form of eq.(\ref{eos2}) can be directly borrowed from the van der waals equation, and only $t<1$ is speculated from our previous works. Eq.(\ref{eos2}) can be named ``the statistical matter model''.

Now we will analyze the effect of eq.(\ref{eos2}) on the universe's evolution. Our model only modifies the pressure of matter, so the cosmic history before the matter-dominated era is believed to be littlely changed and here we only consider the later time. From observations we know that the unverse is always expanding, so $\rho_m$ is decreasing with time. When $\rho_m$ is large enough, $p_m$ is positive but is smaller than eq.(\ref{class-p}). Besides, there exists an era when $p_m\sim0$ like $\Lambda CDM$. 
But it is evident that the pressure in eq.(\ref{eos2}) can be negative If
\begin{equation}
\label{neg-p}
(\frac{\rho_m}{\rho_0})^{1-t}<\frac{\Omega_{x0} }{v^2},
\end{equation}
which tells us that when the matter reaches the thermodynamical equilibrium, its pressure will be negative if its density is low enough. This really is a surprising thing and need to be further confirmed, but it may be natural in our model. If the matter's density is always decreasing, it will always can satisfy eq.(\ref{neg-p}), then can this negative pressure cause the Unverse's accelerating its expansion? From eq.(\ref{friedmann}) this require that $p<-\rho/3$, i.e.
\begin{equation}
(\frac{\rho_m}{\rho_0})^{1-t}<\frac{3\Omega_{x0}}{3v^2+1}.
\end{equation}
Of course this equation will also can be satisfied because of the expansion of the universe, so we find that if the long-range statistical mechanics is really different from the classical one and can produce an EoS like eq.(\ref{eos2}), it will be the matter that drives the universe's accelerated expansion! The next question is that how the universe will evolve in the future, which requires some detailed calculations. We substitute eq.(\ref{eos2}) into the eq.(\ref{conserve}), then the matter's density will evolve as
\begin{eqnarray}
\label{rho-a}
\frac{\rho_m}{\rho_0}&=&[\frac{C(\frac{a_0}{a})^{3B(1-t)}+\Omega_{x0}}{B}]^{\frac{1}{1-t}}\nn\\
&=&[(\frac{C}{B})^{1-t}(\frac{a_0}{a})^{3B}+...+(\frac{\Omega_{x0}}{B})^{\frac{1}{1-t}}]
\end{eqnarray}
where
\begin{equation}
B=1+\frac{v^2}{3}, C=B(1-\Omega_{r0})^{1-t}-\Omega_{x0},
\end{equation}
and $\Omega_r$ is the current radiation's density parameter. Notice that $1<B<4/3$. From observations we know $\kappa\sim0$ and $\Omega_{r0}\sim0.04$, if we define
\begin{equation}
\Omega_m=\frac{\rho_m}{\rho_0},
\end{equation}
its current value $\Omega_{m0}=1-\Omega_{r0}$ will be much closer to 1, which is a major difference of our model from $\Lambda CDM$. When $\Omega_{x0}=0$ (or $a$ is small enough so that $\Omega_{x0}$ can be neglected) and $B$ tends to be $1$, eq.(\ref{rho-a}) will become $\rho\propto a^{-3}$. If $B=1$ and $t=0$, the form of eq.(\ref{rho-a}) will be the same as a direct addition of the matter and DE, so our model includes the result of $\Lambda CDM$, but we still provide a different understanding of the universe and any deviations from $B=1$ or $t=0$ will indicate the preferences of our model. Because of the universe's expansion, from eq.(\ref{rho-a}) we can easily find that the density will finally approximately become a constant
\begin{equation}
\label{rho-min}
\rho_x=\rho_0(\frac{\Omega_{x0}}{B})^{\frac{1}{1-t}},
\end{equation}
If we neglect the variations of $v$ and $\Omega_{x0}$ with time. So our model predicts almost the same destiny of the universe as $\Lambda CDM$: the standard model says that the energy consists of the radiation, the matter and the DE, with the expansion of the universe the DE will dominate and the expansion will accelerate; while our result states that because there is some unknown physics such as long-range statistical mechanics, there is a no-zero minimum of the energy density of the matter even the universe will always expand, which also accelerates the expansion of the universe. However, this minimized density can be smaller than the DE, because: from eq.(\ref{friedmann}) we know that $\ddot{a}>0$ requires that $\rho$ decreases slower than $a^{-2}$, so from eq.(\ref{rho-a}) we know that not only the constant term but also other terms whose power indexes are larger than -2 make the expansion accelerate. So compared with DE, the value of eq.(\ref{rho-min}) should be smaller if $t>0$. Then if in the future the universe's density becomes a constant, according to eq.(\ref{friedmann}) our results may give a smaller value of the Hubble parameter than the standard model, which also differentiates our results from the standard model.

Before analyzing the data of SNIa, we need to calculate the Hubble parameter. It can be expressed by four parameters $(t, \Omega_{x0}, B, h_0)$ in our model,
\begin{eqnarray}
\frac{H(z)}{H_0}&=&(\frac{C(1+z)^{-3B(1-t)}+\Omega_{x0}}{B})^{\frac{1}{1-t}}\nn\\
&&+\Omega_{r0}(1+z)^4.
\end{eqnarray}
where $h_0$ is the present value of the Hubble parameter $H_0$ in unit $100km/s/Mpc$ and $z$ is the redshift.
\section{Compared with the data of SNIa}
\label{snIa}
\subsection{Data Analysis}
To check our proposals made in Sec. II and to constrain the model
parameters, we take use of the Union2 dataset~\cite{Union2} and the Hubble evolution data.

The Union2 dataset contains 557 type Ia SN data and covers the
redshift range $z=[0.015,1.4]$, including samples from other
surveys, such as CfA3~\cite{cfa}, SDSS-II Supernova
Search~\cite{sdss} and high-z Hubble Space Telescope.

We fit the SNIa data by minimizing the $\chi^2$ value of the
distance modulus. $\chi_{sn}^{2}$ for SNIa is obtained by comparing
theoretical distance modulus
$\mu_{th}(z)=5\log_{10}[(1+z)\int_{0}^{z}dx/E(x)]+\mu_{0}$
($\mu_{0}=42.384-5\log_{10}h_0, E(z)=H(z)/H_z$) with observed
$\mu_{ob}$ of supernova:
\[
\chi_{sn}^{2}=\sum_{i}^{557}\frac{[\mu_{th}(z_{i})-\mu_{ob}(z_{i})]^{2}}{\sigma^{2}(z_{i})}.
\]

To reduce the effect of $\mu_{0}$, we expand $\chi_{sn}^{2}$ with
respect to $\mu_{0}$ \cite{Nesseris:2005ur}:
\begin{equation}
\chi_{sn}^{2}=F+2G\mu_{0}+H\mu_{0}^{2}\label{eq:expand}\end{equation}
 where \begin{eqnarray*}
F & = & \sum_{i}\frac{[\mu_{th}(z_{i};\mu_{0}=0)-\mu_{ob}(z_{i})]^{2}}{\sigma^{2}(z_{i})},\\
G & = & \sum_{i}\frac{\mu_{th}(z_{i};\mu_{0}=0)-\mu_{ob}(z_{i})}{\sigma^{2}(z_{i})},\\
H & = & \sum_{i}\frac{1}{\sigma^{2}(z_{i})}\end{eqnarray*}
 (\ref{eq:expand}) has a minimum as \[
\widetilde{\chi}_{sn}^{2}=\chi_{sn,min}^{2}=F-G^{2}/H\]
 which is independent of $\mu_{0}$. In fact, it is equivalent to
performing an uniform marginalization over $\mu_{0}$, the difference
between $\widetilde{\chi}_{sn}^{2}$ and the marginalized
$\chi_{sn}^{2}$ is just a constant \cite{Nesseris:2005ur}. We will
adopt $\widetilde{\chi}_{sn}^{2}$ as the goodness of fit between
theoretical model and SNIa data.

We also use the $12$ Hubble evolution data from \cite{Simon:2004tf}
and \cite{Gaztanaga:2008xz}, its $\chi_{H}^{2}$ is defined as

\[
\chi_{H}^{2}=\sum_{i=1}^{12}\frac{[H(z_{i})-H_{ob}(z_{i})]^{2}}{\sigma_{i}^{2}}.\]
 Note that the redshift of these data falls in the region $z\in(0,1.75)$.

In summary, \[
\chi_{tot}^{2}=\widetilde{\chi}_{sn}^{2}\chi_{H}^{2}\] and we assume
uniform priors on all the parameters. We also assumed the prior that
the age of our universe $T_0$ satisfies $10Gyr<T_0<20Gyr$.

\subsection{Results and Discussions}

\begin{figure}
\begin{centering}
\includegraphics[width=0.5\textwidth]{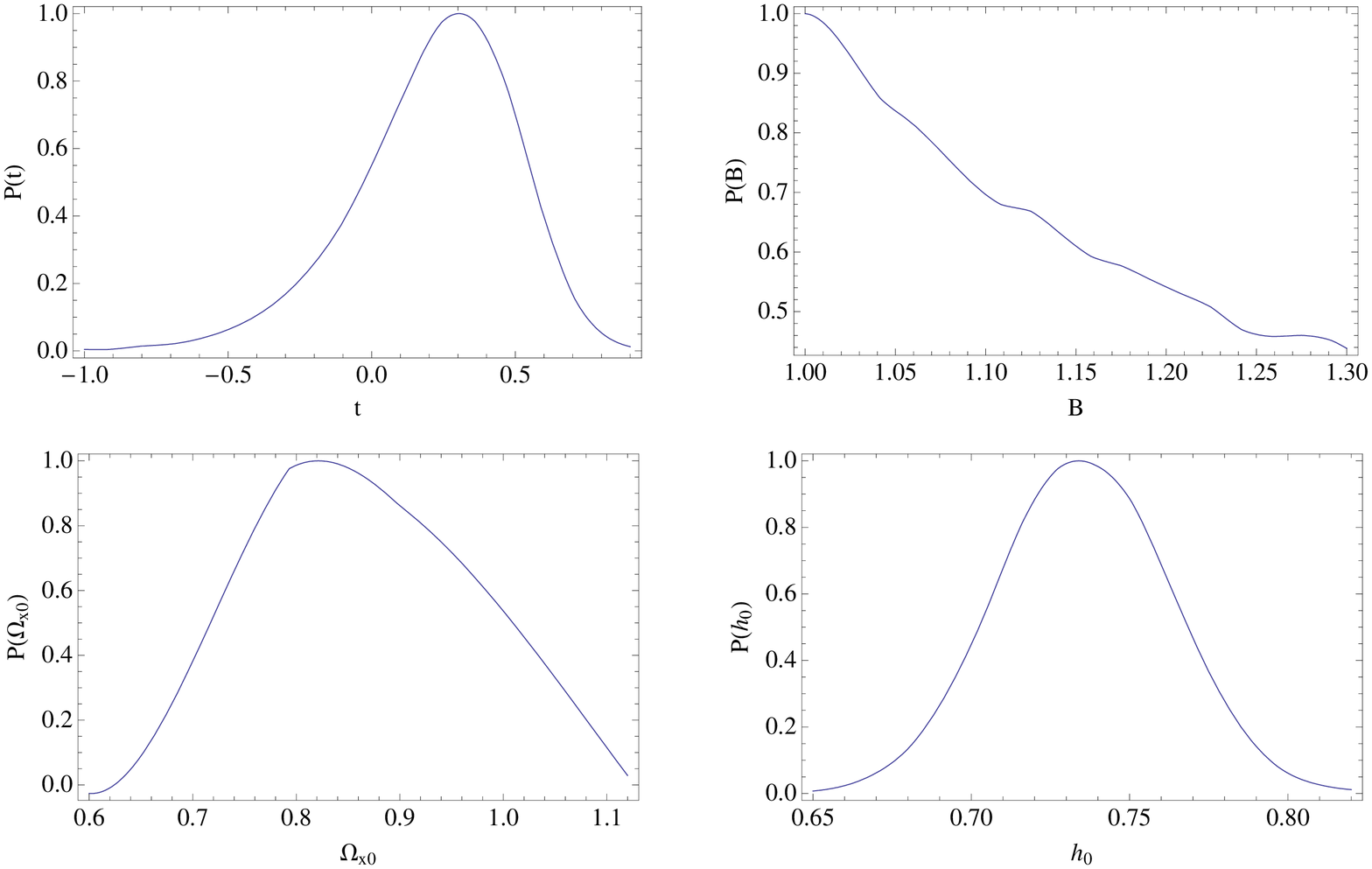}
\par\end{centering}
\caption{\label{fig:likelihood}1D marginalized distribution
probability of $t,\,\Omega_{x0},\,B,\,h_0$.}
\end{figure}

The analysis is performed by using the Monte Carlo Markov Chain in
the multidimensional parameter space to derive the likelihood.
Naturally, we employ some physically obvious limitations to make the
estimation of parameters more robust, for example, we set $t<1,
1\leq B<4/3$ in our analysis. We first investigate the constraint on
the model parameters, and the best-fit values and errors of
parameters are summary in Tab.\ref{tab:fit_result}. We also plot the 1D marginalized distribution
probability of each parameter, shown in Fig.\ref{fig:likelihood}.
The likelihood distribution shows a remarkable deviation from the
Gaussian distribution, which result in a discrepancy between the
best-fit values and the expected values of the parameters, as
consistent with the result in Tab.\ref{tab:fit_result}.

\begin{table}
\begin{centering}
\begin{tabular}{|c|c|c|c|}
\hline$t$  & $\Omega_{x0}$  & $B$  & $h_0$\tabularnewline \hline
 $0.237_{+0.346,\,+0.562}^{-0.347,\,-1.237}$  &
$0.824_{+0.200,\,+0.271}^{-0.107,\,-0.161}$  & $1.030$ &
$0.724_{+0.050,\,+0.074}^{-0.025,\,-0.052}$\tabularnewline
\hline  $0.051$  & $0.734$  & $1.008$ & $0.738$\tabularnewline
\hline
\end{tabular}
\par\end{centering}

\caption{\label{tab:fit_result}Expected values, $1-\sigma$ and
$2-\sigma$ error of $t, \Omega_{x0}, B, h_0$ in this model. The
second line is in form of
$\text{Expectation}_{+1\sigma,\,+2\sigma}^{-1\sigma,\,-2\sigma}$.
And the best-fit values are also shown in the last line, which are
different from the expected values.}

\end{table}

\begin{figure}
\begin{centering}
\includegraphics[height=0.35\textwidth,width=0.35\textwidth]{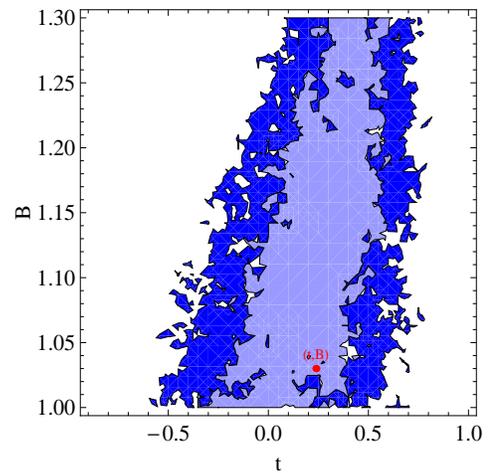}
\par\end{centering}
\caption{\label{fig:contour}$68\%$ and $95\%$ contour plot in $t-B$
plane. Red dot in the center is the expected value.}
\end{figure}
Further, we focus on parameters $(t, B)$, which play a more
important role in our model. The 2D contour plot is shown in
Fig.\ref{fig:contour}, which shows that the results of standard $\Lambda CDM$
model is contained in our model. The $2\sigma$ errors mean
that the case with $t>0.80$ is excluded at $95.4\%$ confidence
level, which is consistent with our assumption. But constraint
on $B$ can not be obtained.

From above results, we find that the data of SNIa can not differentiate our model from $\Lambda CDM$. We need to consider the perturbation theory to further test our model, which will be another work. But from Table.\ref{tab:fit_result} and Figure.\ref{fig:likelihood} we can see that the data seem to indicate some deviations from $B=1$ and $t=0$, i.e. $\Lambda CDM$. Notice that according to current criteria, the dark matter is called to be cold if $v<0.1$\footnote{see \href{Wikipedia}{http://en.wikipedia.org/wiki/Dark_matter}}, which means that $B<1.0033$, so there is only a very short range of $B$ for the cold matter. however, both the best-fit and expected value of $B$ are larger than $1.0033$, which suggests that the matter may be not so cold as $\Lambda CDM$ predicts. This conclusion also agrees with some works \cite{Colin:2000dn,Bode:2000gq} which aim to use the warm dark matter to solve the contradictions between observations and numerical simulations of $\Lambda CDM$ at the galaxy scale.

\section{Conclusion}
Our previous works have proved that the thermodynamics of
gravitating systems at the galaxy scale may be different from the
classical one, and we obtain an EoS like eq.(\ref{eos3}) which is in
analogy with the van der waals equation. So we speculate that it is
the long-range statistical mechanics that produces the
eq.(\ref{eos3}). Based on possible effect of the general relativity
on the matter's EoS, we propose that at the cosmological scale the
form of the matter's EoS may be like the eq.(\ref{eos2}). Then we
find that with the universe's expansion the density will decrease
with time to be a no-zero minimized constant matter density, and it
is the matter that drives the expansion's acceleration. So our model
provides a new kind of explanation for the expansion's acceleration, but
there are three major different results of our model from $\Lambda CDM$:

(1). $\Omega_m=1-\Omega_r$ is close to 1 and is much larger than its value in $\Lambda CDM$;

(2). $B$ and $t$ may be not exactly to $1$ and $0$ respectively;

(3). In the future when the unverse's density becomes a constant, its value can be smaller than DE.

It is more easy to compare our model with $\Lambda CDM$ by the second point,
so we use the data of SNIa to constrain the value of $B$ and $t$. From the
results we can not differentiate our model from $\Lambda CDM$, but
there seems some indications of deviations from $\Lambda CDM$, especially we
find that the matter may be not so cold as previously believed. To
further test our model, we need to consider the density's perturbation in
the future.

\begin{acknowledgments}

DBK is very grateful for Zhong-Liang Tuo's many helps of dealing with the data of supernova. This work is supported by the National Basic Research Programm of China, NO:2010CB832805.

\end{acknowledgments}

\end{document}